\documentclass[journal]{IEEEtran}

\usepackage{cite} 

\usepackage[caption=false,font=footnotesize]{subfig}	
\usepackage{graphicx}
\usepackage{psfrag}

\usepackage{algorithm}
\usepackage{algorithmicx}
\usepackage{algpseudocode}

\usepackage{amsmath,amsfonts,amssymb,amsthm}
\usepackage{mathtools}
\usepackage{mathrsfs}	
\usepackage{subdepth}	

\usepackage{array}	
\usepackage{tabularx}   

\usepackage{enumitem}

\usepackage[hidelinks]{hyperref}
\hypersetup{
 pdftitle={},%
 pdfauthor={},%
 pdfsubject={},%
 pdfkeywords={},%
 colorlinks={},%
 linkcolor={blue},%
 linktocpage={},%
 pageanchor={},
 citecolor={}
 }
\usepackage[sort,compress,capitalise]{cleveref}	
\usepackage{url}

\usepackage[dvipsnames]{xcolor} 
\usepackage{xspace} 

\usepackage{tikz}
\usetikzlibrary{babel}  
\usepackage[straightvoltages,nooldvoltagedirection]{circuitikz}

{
\theoremstyle{plain}
\newtheorem{Hypothesis}{Hypothesis}

\newtheorem{Theorem}{Theorem}

}

{
\theoremstyle{definition}

}

\crefname{Hypothesis}{Hyp.}{Hyps.}
\Crefname{Hypothesis}{Hyp.}{Hyps.}

\crefname{Lemma}{Lemma}{Lemmata}
\Crefname{Lemma}{Lemma}{Lemmata}

\crefname{Theorem}{Theorem}{Theorems}
\Crefname{Theorem}{Theorem}{Theorems}

\crefname{Definition}{Def.}{Defs.}
\Crefname{Definition}{Def.}{Defs.}

\newtheorem{remark}{Remark}
\newtheorem{condition}{Condition}
\newtheorem{definition}{Definition}

\newcommand{\CIGRE}{CIGR{\'E}\xspace}

\newcommand{\DFT}[1][]{DFT\xspace}  

\newcommand{\LTP}{LTP\xspace}	

\newcommand{\TE}[1][]{\textup{TE#1}\xspace}   
\newcommand{\NE}[1][]{\textup{NE#1}\xspace}   

\newcommand{\DER}[1][]{DER#1\xspace}    
\newcommand{\CIDER}[1][]{CI\DER[#1]}    

\newcommand{\TDS}{TDS\xspace}	
\newcommand{\HPF}{HPF\xspace}	
\newcommand{\PF}{PF\xspace}	    
\newcommand{\NR}{NR\xspace}	    

\newcommand{\CompNum}{\mathbb{C}}   
\newcommand{\RelNum}{\mathbb{Z}}   
\newcommand{\Real}[1]{\Re\left\{#1\right\}}

\newcommand{\Abs}[1]{\left|#1\right|}

\newcommand{\Set}[1]{\mathcal{#1}}
\newcommand{\Card}[1]{\left|#1\right|}

\newcommand{\diag}{\operatorname{diag}}

\newcommand{\col}{\operatorname{col}}

\newcommand{\Graph}[1]{\mathfrak{#1}}

\newcommand{\grid}{\gamma}
\newcommand{\pwr}{\pi}
\newcommand{\ctrl}{\kappa}
\newcommand{\act}{\alpha}

\newcommand{\trafo}{\tau}
\newcommand{\refr}{\rho}
\newcommand{\spt}{\sigma}

\newcommand{\V}{\mathbf{V}}	
\newcommand{\I}{\mathbf{I}}	
\newcommand{\Z}{\mathbf{Z}}     
\newcommand{\Y}{\mathbf{Y}}     
\newcommand{\HG}{\mathbf{H}}    
\newcommand{\G}{\mathbf{G}}     
\newcommand{\W}{\mathbf{W}}     

\newcommand{\ground}{\Set{G}}

\newcommand{\nodes}{\Set{N}}
\newcommand{\branches}{\Set{L}}
\newcommand{\shunts}{\Set{T}}

\newcommand{\branchgraph}{\Graph{B}}

\newcommand{\harmonics}{\Set{H}}
\newcommand{\formers}{\Set{S}}
\newcommand{\followers}{\Set{R}}
\newcommand{\passiveloads}{\Set{P}}

\newcommand{\PermutationFormer}{\mathbf{P}_{\formers}}
\newcommand{\PermutationFollower}{\mathbf{P}_{\followers}}

\newcommand{\RMS}{RMS\xspace}

\begin{document}

\title{\huge{%
    Ensuring Solution Uniqueness in Fixed-Point-Based Harmonic Power Flow Analysis with Converter-Interfaced Resources: Ex-post Conditions
}}

\author{%
    Antonio Di Pasquale, Johanna~Kristin~Maria~Becker,~\IEEEmembership{Member,~IEEE}, \\	Andreas~Martin~Kettner,~\IEEEmembership{Member,~IEEE},
and~Mario~Paolone,~\IEEEmembership{Fellow,~IEEE}%
  \thanks{A. Di Pasquale is with the Department of Electrical Engineering and Information Technology at the University of Naples Federico II (E-mail: antonio.dipasquale@unina.it).}
 \thanks{J. Becker and M. Paolone are with the Distributed Electrical Systems Laboratory at the {\'E}cole Polytechnique F{\'e}d{\'e}rale de Lausanne (EPFL) in CH-1015 Lausanne, Switzerland (E-mail: \{johanna.becker, mario.paolone\}@epfl.ch).}
	\thanks{A. Kettner is with PSI NEPLAN AG, 8700 Küsnacht, Switzerland (E-mail: andreas.kettner@neplan.ch).}%
	\thanks{This work was funded by the Schweizerischer Nationalfonds (SNF, Swiss National Science Foundation) via the National Research Programme NRP~70 ``Energy Turnaround'' NRP 70 ``Energy Turnaround'' (projects nr. 197060).}%
}

\maketitle






















\begin{abstract}
    Recently, the authors of this paper proposed a method for the Harmonic Power-Flow (HPF) calculus in polyphase grids with widespread deployment of Converter-Interfaced Distributed Energy Resources (\CIDER[s]). The \HPF problem was formulated by integrating the hybrid nodal equations of the grid with a detailed representation of the \CIDER[s] hardware, sensing, and controls as \textit{Linear Time-Periodic} (\LTP) systems, and solving the resulting mismatch equations using the Newton-Raphson (NR) method. This work introduces a novel problem formulation based on the fixed-point algorithm that, combined with the contraction property of the \HPF problem, provides insights into the uniqueness of its solution. Notably, the effectiveness of the fixed-point formulation and the uniqueness of the solution are evaluated through numerical analyses conducted on a modified version of the \CIGRE low-voltage benchmark microgrid.
\end{abstract}
\section*{Nomenclature}

\begin{center}
    
\begin{tabularx}{\columnwidth}{p{3cm}p{5cm}}
    \multicolumn{2}{c}{\textit{\HPF Study}}\\
\end{tabularx}
\begin{IEEEdescription}[\IEEEusemathlabelsep\IEEEsetlabelwidth{$\formers\cup\followers$}]
    \item[$\formers$]
        The nodes with grid-forming \CIDER[s]
    \item[$\followers$] 
        The nodes with grid-following \CIDER[s]
        \item[$\mathbf{P}$] 
        The permutation matrix
\end{IEEEdescription}

\end{center}

\begin{center}
    
\begin{tabularx}{\columnwidth}{p{3cm}p{5cm}}
    \multicolumn{2}{c}{\textit{Grid Model}}\\
\end{tabularx}
\begin{IEEEdescription}[\IEEEusemathlabelsep\IEEEsetlabelwidth{$\formers\cup\followers$}]
    \item[$n\in\nodes$]
        A three-phase node ($\nodes\coloneqq\{1,...,N\}$)
    \item[$\V_{n}$] 
        The phasors of the phase-to-ground nodal voltages at node $n\in\nodes$
    \item[$\I_{n}$]
        The phasors of the injected currents at node $n\in\nodes$
    \item[$\ell\in\branches$] 
        A branch element ($\ell=(m,n):~m,n\in\nodes$)
    \item[$\Z_{\ell}$] 
        A compound branch impedance at $\ell\in\branches$
    \item[$t\in\shunts$] 
        A shunt element ($t=(n,g):~n\in\nodes,~g\in\ground$)
    \item[$\Y_{t}$] 
        A compound shunt admittance at $t\in\shunts$
    \item[$\mathbf{A}_{\branchgraph}$] 
        The three-phase branch incidence matrix
    \item[$\Y$]
        The compound nodal admittance matrix
    \item[$\I_{\formers}$]
        The phasors of the injected currents at all $s\in\formers$
    \item[$\V_{\followers}$]
        The phasors of the phase-to-ground nodal voltages at all $r\in\followers$
    \item[$\Y_{\formers\times\followers}$] 
        The block of $\Y$ linking $\I_{\formers}$ and $\V_{\followers}$
    \item[$\HG$] 
        The compound nodal hybrid matrix (w.r.t. $\formers,\followers$)
    \item[$\HG_{\formers\times\followers}$] 
        The block of $\HG$ linking $\V_{\formers}$ and $\V_{\followers}$
    \item[$f$] 
        An arbitrary frequency
    \item[$f_1$]
        The fundamental frequency ($f_{1}\coloneqq\frac{1}{T}$)
    \item[$h\in\harmonics$] 
        A harmonic order ($\harmonics\coloneqq\{-h_{\max},\ldots,h_{\max}\}$)
    \item[$f_{h}$]
        The harmonic frequency of order $h$ ($f_{h}\coloneqq h\cdot f_{1}$)
    \item[$\hat{\V}_{\formers}$] 
        The column vector composed of the Fourier coefficients of $\V_{\formers}$
    \item[$\hat{\HG}_{\formers\times\followers}$] 
        The Toeplitz matrix of the Fourier coefficients of $\HG_{\formers\times\followers}$ (i.e., $\HG_{\formers\times\followers}(f)$ evaluated at $f=f_{h}$)
\end{IEEEdescription}

\end{center}

\begin{center}
    
\begin{tabularx}{\columnwidth}{p{3cm}p{5cm}}
    \multicolumn{2}{c}{\textit{\CIDER Model}}\\
\end{tabularx}
\begin{IEEEdescription}[\IEEEusemathlabelsep\IEEEsetlabelwidth{$\formers\cup\followers$}]
    \item[$\grid$]
        The power grid
    \item[$\pwr$]
        The power hardware of a \CIDER
    \item[$\ctrl$]
        The control software of a \CIDER
    \item[$\act$]
        The actuator of a \CIDER
    \item[$\refr$] 
        The reference calculation of a \CIDER
    \item[$\spt$] 
        The setpoint of a \CIDER
    \item[$\lambda$]
        A generic stage inside the cascaded structure of a \CIDER ($\lambda\in\{1,\ldots,\Lambda\}$)
    \item[$\varphi_{\lambda}$] 
        The filter element associated with stage $\lambda$
    \item[$\ctrl_{\lambda}$] 
        The controller element associated with stage $\lambda$
    \item[$\mathbf{x}(t)$]
        The state vector of a state-space model
    \item[$\mathbf{u}(t)$]
        The input vector of a state-space model
    \item[$\mathbf{y}(t)$]
        The output vector of a state-space model
    \item[$\mathbf{w}(t)$]
        The disturbance vector of a state-space model
    \item[$\mathbf{A}(t)$]
        The system matrix of an \LTP system
    \item[$\mathbf{B}(t)$]
        The input matrix of an \LTP system
    \item[$\mathbf{C}(t)$]
        The output matrix of an \LTP system
    \item[$\mathbf{D}(t)$]
        The feed-through matrix of an \LTP system
    \item[$\mathbf{E}(t)$]
        The input disturbance matrix of an \LTP system
    \item[$\mathbf{F}(t)$]
        The output disturbance matrix of an \LTP system
    \item[$\trafo_{\ctrl|\pwr}$]
        A change of reference frame from $\pwr$ to $\ctrl$
    \item[$\mathbf{T}_{\ctrl|\pwr}(t)$] 
        The \LTP matrix which describes $\trafo_{\ctrl|\pwr}$
   \item[$\mathbf{X}_{h}$] 
        The Fourier coefficients of $\mathbf{x}(t)$ ($h\in\harmonics$)
    \item[$\hat{\mathbf{X}}$] 
        The column vector composed of the $\mathbf{X}_{h}$
    \item[$\mathbf{A}_{h}$] 
        The Fourier coefficients of $\mathbf{A}(t)$ ($h\in\harmonics$)
    \item[$\hat{\mathbf{A}}$] 
        The Toeplitz matrix composed of the $\mathbf{A}_{h}$
    \item[$\Hat{\mathbf{G}}$] 
        The harmonic-domain closed-loop gain
\end{IEEEdescription}

\end{center}
\section{Introduction}\label{sec:Introduction}
As acknowledged by existing literature, the extensive deployment of \textit{Converter-Interfaced Distributed Energy Resources} (\CIDER[s]) in modern power systems poses potential risks to grid operation. As highlighted in~\cite{wang2018harmonic}, the widespread integration of power electronic-based systems introduces new challenges to such systems' stability and power quality. Electronic power converters' broad timescale and frequency-coupling dynamics often result in harmonic instability, manifesting as resonances or abnormal harmonics across a wide frequency range. In this context, the authors of this work recently introduced in~\cite{kettner2021harmonic_part_I} and \cite{becker2021harmonic_part_II} a \textit{Harmonic Power-Flow} (\HPF) method designed for polyphase grids in the presence of \CIDER[s]. Within this framework, the grid is modelled using the polyphase circuit theory, and the \CIDER[s] are represented by \textit{Linear Time-Periodic} (\LTP) models. Notably,  the \HPF relies on the nodal equations, which are derived from the hybrid equations from the grid's point of view, and the closed-loop transfer functions from the point of view of the \CIDER[s]. Specifically, the latter characterize the behaviour of the \CIDER[s] regarding the generation and propagation of harmonics resulting from interactions among power hardware, control software, and reference calculation. Particularly, combining grid and \CIDER[s] equations yields a system of mismatch equations, which must be equal to zero at equilibrium. Due to its nonlinearity, the problem is solved using the \emph{Newton-Raphson} (NR) method. It is important to note that the harmonic stability is inherently related to the solvability of the \HPF problem. By definition, a system is deemed unstable when the equations describing its behaviour lack equilibrium points~\cite{8750828,kundur2004definition}. 
Within the context of classical \emph{Power-Flow} (\PF) analysis, the unsolvability of power-flow equations indicates that the system cannot attain a steady-state equilibrium. 
Moreover, in terms of the solvability of conventional \PF, it is widely acknowledged that, due to the nonlinear nature of the equations, the existence of real-valued solutions is not guaranteed. 
Networks with light loading often have multiple solutions~\cite{molzahn2016toward}, while a sufficiently loaded network may have none. 
However, in cases with multiple solutions, there is typically only one feasible solution that is characterized by high-voltage magnitudes (i.e., close to 1~pu) at buses and small branch current flows, referred to as stable~\cite{simpson2017theory}.

Thus, the present work aims to extend these considerations to the aforementioned \HPF study. In pursuit of this objective, the fixed-point method has been identified as the most suitable approach to investigate the solvability of the \HPF problem.
More precisely, based on the contraction property, one can establish conditions for the uniqueness of the solution. The primary contributions of this paper are as follows:
\begin{itemize}
    \item Development and validation of an alternative formulation for the \HPF study proposed in~\cite{kettner2021harmonic_part_I,becker2021harmonic_part_II} based on the fixed-point method.
    \item Provision of ex-post conditions derived from the contraction property of the new \HPF formulation to prove the uniqueness of the solution.
\end{itemize}

The remainder of this paper is organized as follows. Section~\ref{Sec:RelatedLiterature} provides an overview of the fixed-point formulation for the conventional \PF problem, along with a discussion on how the existing literature addresses the uniqueness of the solution to the \PF problem.  Section~\ref{sec:HarmonicPowerFlow} recalls the main theoretical foundation of the \HPF framework introduced by the authors in the previous works. Subsequently, Section~\ref{sec:Fixed-PointFormulation} derives the new fixed-point formulation of the \HPF study and the ex-post conditions guaranteeing the uniqueness of the solution. Section~\ref{sec:NumericalValidation} conducts a numerical investigation on a modified version of the \CIGRE low-voltage benchmark microgrid to assess the applicability of the fixed-point \HPF and examines the conditions for the ex-post evaluation of solution uniqueness. Finally, Section~\ref{sec:Conclusion} draws the main conclusions.
\section{Related Literature}\label{Sec:RelatedLiterature}
This section does not aim to provide an exhaustive literature review; rather, its purpose is to discuss the works that the authors consider most pertinent in the context of the present study. 
For a more extensive review of the existing literature, readers are directed to~\cite{molzahn2013sufficient,bernstein2018load}.

In the last decades, considerable attention has been focused on the solvability of the \PF equations and the quest for its unique solutions. Iterative methods, such as the \NR algorithm, are commonly employed. However, the latter exhibits high sensitivity to initialization~\cite{tinney1967power}, making it challenging to distinguish between poor initialization versus problem infeasibility in the case of non-convergence. 
To tackle this challenge, studies employing fixed-point, or contraction-based approaches to investigate the solvability of the \PF problem were proposed. 
These approaches provide sufficient conditions for the existence, and often the uniqueness, of a suitable \PF solution. Specifically, the contraction property of the problem inherently drives a fixed-point iteration to compute this solution, ensuring convergence from any initialization point within the contraction region~\cite{chen2023fixed}.

Notably, the first attempts in applying the fixed-point techniques to \PF studies can be traced back to~\cite{meisel1970application}, which primarily delves into the convergence property of the \NR method. 
A more recent development is presented in~\cite{lisboa2014fast}, introducing an effective fixed-point \PF method tailored for radial distribution networks. However, it lacks a comprehensive discussion on the convergence and solvability of the problem. 
In this regard, a noteworthy contribution is made by~\cite{bolognani2015existence}, which establishes sufficient conditions for the existence and uniqueness of the \PF solution in balanced distribution grids. 
\cite{yu2015simple} extends the criteria proposed in~\cite{bolognani2015existence} and develops a new mathematical criterion for the certification of solvability. A further step forward is made by~\cite{wang2016explicit}, which applies the Banach-Caccioppoli theorem for balanced radial or meshed distribution networks.  This approach, in comparison to prior works, introduces less stringent sufficient conditions to guarantee the existence and uniqueness of solutions as a function of the system state. This analysis was extended to unbalanced networks in \cite{wang2017existence,bernstein2018load} 

Recently, a novel \PF formulation for electrical distribution systems was proposed by~\cite{giraldo2022fixed}, employing the current injection method and utilizing the Laurent series expansion. The convergence analysis of the algorithm is established through the Banach-Caccioppoli fixed-point theorem, ensuring both numerical stability and uniqueness of the solution, irrespective of the initialization.

In this respect, the present work draws inspiration from the existing literature on the solvability and uniqueness of solutions in conventional \PF problems. It aims to develop a fixed-point formulation for the \HPF study and evaluate the uniqueness of its solution based on the contraction property.

\section{Harmonic Power-Flow mondel in the Presence of Converter-Interfaced Resources }\label{sec:HarmonicPowerFlow}

In~\cite{becker2021harmonic_part_II}, it is observed that the \NR algorithm employed for solving the \HPF study consistently converges to the same solution, apparently irrespective of the chosen initial point. 
This observation suggests that the \HPF method may possess specific mathematical properties ensuring the uniqueness of its solution. 
 Hence, as mentioned in Section~\ref{sec:Introduction}, this paper delves into the analysis of the mathematical properties of the \HPF framework proposed in \cite{kettner2021harmonic_part_I,becker2021harmonic_part_II}, with a specific focus on identifying ex-post conditions necessary for ensuring uniqueness of the solution. 
In this regard, the present section revisits the key theoretical fundamentals proposed in those works for modelling both the grid and the \CIDER[s]. 
Finally, the mathematical formulation describing the \HPF problem is recalled.

\subsection{Model of the Grid}
In~\cite{kettner2021harmonic_part_I}, the authors considered a generic three-phase grid equipped with an effectively grounded neutral conductor. The grid is represented by lumped elements, which are categorized into branches $\ell \in \mathcal{L}$, associated with the \textit{compound branch impedance matrix $\Z_{\ell}$},  and shunts $t \in \mathcal{T}$, associated with the \textit{compound shunt admittance matrix} $\Y_{t}$.

In order to develop the model of the grid, the authors have made the following assumptions:
\begin{Hypothesis}\label{hyp:grid:ground}
    The grid is equipped with a neutral conductor which is effectively grounded. As a result, the voltage between the neutral conductor and the physical earth is negligible. 
\end{Hypothesis}
\begin{Hypothesis}\label{hyp:grid:model}
The components of the grid are linear and passive. Additionally, there is no electromagnetic coupling among them. 
\end{Hypothesis}
Indeed, electromagnetic coupling only matters within each component.
As a consequence of \cref{hyp:grid:model}, the circuit equations referring to the grid can be formulated independently at each frequency $f$ using either impedance or admittance parameters.
\begin{Hypothesis}\label{hyp:grid:parameters}
	The compound branch impedance matrices $\Z_{\ell}$ are symmetric, invertible, and lossy at all frequencies:
	\begin{equation}
		\Z_{\ell}(f):
		\quad
		\left[~
		\begin{aligned}
			&\Z_{\ell}(f)= (\Z_{\ell}(f))^{T}	\\
			&\exists\Y_{\ell}(f)= (\Z_{\ell}(f))^{-1}	\\
			&\Real{\Z_{\ell}(f)}\succeq 0
		\end{aligned}
		\right.
		\label{eq:branches:parameters}
	\end{equation}
	The compound shunt admittance matrices $\Y_{t}$ are symmetric, invertible, and lossy at all frequencies if they are nonzero:
	\begin{equation}
		\text{if}~\Y_{t}(f)\neq\mathbf{0}:
		\quad
		\left[~
		\begin{aligned}
			&\Y_{t}(f) =          (\Y_{t}(f))^{T}\\
			&\exists\Z_{t}(f) =          (\Y_{t}(f))^{-1}\\
			&\Real{\Y_{t}(f)} \succeq	0
		\end{aligned}
		\right.
		\label{eq:shunts:parameters}
	\end{equation}
\end{Hypothesis}
Let $\mathcal{N}$ represent the set of all nodes.
Then, based on the hypotheses reported above, the \textit{compound nodal admittance matrix} $\Y(f) \in \mathbb{C}^{3|\mathcal{N}|\times 3|\mathcal{N} |}$ is defined, at each frequency $f$,  as follows: 
\begin{equation}
    \Y(f)=\mathbf{A}^{T}_{\branchgraph}\Y_{\branches}(f)\mathbf{A}_{\branchgraph} + \Y_{\shunts}(f)
    \label{eq:nodes:admittance}
\end{equation}
where $(\cdot)^{T}$ is the transposed matrix, $\mathbf{A}_{\branchgraph}\in\RelNum^{3\Card{\branches}\times3\Card{\nodes}}$ is the three-phase incidence matrix, whereas $\Y_{\branches}(f) \in~   \CompNum^{3\Card{\branches}\times3\Card{\branches}}$ and $\Y_{\shunts}(f)\in\CompNum^{3\Card{\nodes}\times3\Card{\nodes}}$ are the \textit{primitive compound admittance matrices} associated with the branches and shunts, respectively~\cite{alvarado1982formation}, whose expressions are:
\begin{alignat}{2}
    \Y_{\branches}(f)
    &\coloneqq          \diag_{\ell\in\branches}(\Y_{\ell}(f))&
    &~\in~   \CompNum^{3\Card{\branches}\times3\Card{\branches}}
    \label{eq:branches:primitive}\\
    \Y_{\shunts}(f)
    &\coloneqq          \diag_{t\in\shunts}(\Y_{t}(f))&
    &~\in~   \CompNum^{3\Card{\nodes}\times3\Card{\nodes}}
    \label{eq:shunts:primitive}
\end{alignat}

The admittance matrix $\Y$ links the nodal injected currents $\I\in\CompNum^{3\Card{\nodes}\times1}$ and the phase-to-ground voltages  $\V\in\CompNum^{3\Card{\nodes}\times1}$ as follows:
\begin{equation}
       \I(f)=\Y(f)\V(f)
    \label{eq:nodes:current}
\end{equation}
However, \cite{kettner2021harmonic_part_I} points out that in the \HPF framework it is more convenient to model the system by the nodal hybrid equations, where the unknown variables are the nodal injected currents $\I_{\formers}(f)$ at the nodes $\formers$ with grid-forming \CIDER[s] and the nodal phase-to-ground voltages $\V_{\followers}(f)$ at the nodes $\followers$ with grid-following \CIDER[s], respectively (see \cref{sec:CiderModel}). Hence, \eqref{eq:nodes:current} is replaced by the following hybrid system of equations~\cite{kettner2018properties}:
\begin{equation}
    	    \begin{bmatrix}
    	        \V_{\formers}(f)\\
    	        \I_{\followers}(f)
    	    \end{bmatrix}
        =   \begin{bmatrix}
    	            \HG_{\formers\times\formers}(f)
                &   \HG_{\formers\times\followers}(f)\\
                    \HG_{\followers\times\formers}(f)
                &   \HG_{\followers\times\followers}(f)
    	    \end{bmatrix}
    	    \begin{bmatrix}
    	        \I_{\formers}(f)\\
    	        \V_{\followers}(f)
    	    \end{bmatrix}
        \label{eq:nodes:hybrid}
\end{equation}
where $\HG_{\formers\times\formers}$, $\HG_{\formers\times\followers}$, $\HG_{\followers\times\formers}$ and $\HG_{\followers\times\followers}$ are the blocks of the \textit{compound nodal hybrid matrix} $\mathbf{H}$ associated to the sets $\formers$ and $\followers$, respectively.  
\begin{remark}
    In~\cite{kettner2018properties}, the authors demonstrated that if~\cref{hyp:grid:ground,hyp:grid:parameters} hold, every proper diagonal sub-block of the compound admittance matrix $\Y$ has full rank. Consequently, the compound nodal hybrid matrix $\mathbf{H}$ exists.
\end{remark}
It is important to note that ~\eqref{eq:nodes:hybrid} refers to a specific frequency. However, this representation can be extended to cover a desired spectrum of harmonic frequencies span $\harmonics$ w.r.t. a given fundamental frequency $f_1$ (i.e., $f_{h} \coloneqq h \cdot f_{1}, ~ h\in\harmonics\subset\mathbb{Z}$) as follows:
\begin{equation}
    \begin{bmatrix}
	        \hat{\V}_{\formers}\\
	        \hat{\I}_{\followers}
	    \end{bmatrix}
    =   \begin{bmatrix}
	            \hat{\HG}_{\formers\times\formers}
            &   \hat{\HG}_{\formers\times\followers}\\
                \hat{\HG}_{\followers\times\formers}
            &   \hat{\HG}_{\followers\times\followers}
	    \end{bmatrix}
	    \begin{bmatrix}
	        \hat{\I}_{\formers}\\
	        \hat{\V}_{\followers}
	    \end{bmatrix}
	   \label{eq:nodes:hybrid:harmonics}
\end{equation}
where
\begin{alignat}{2}
    \hat{\V}_{\formers}
    &\coloneqq          \col_{h\in\harmonics}(\V_{\formers}(f_{h}))&
    &~\in~   \CompNum^{3\Card{\harmonics}\Card{\formers}\times1}\label{eq:Hybrid:Voltage}
    \\
    \hat{\HG}_{\formers\times\formers}
    &\coloneqq          \diag_{h\in\harmonics}(\HG_{\formers\times\formers}(f_{h}))&
    &~\in~   \CompNum^{3\Card{\harmonics}\Card{\formers}\times3\Card{\harmonics}\Card{\formers}}
\end{alignat}
The other block matrices and vectors are defined analogously.

\subsection{Generic Model of the \CIDER[s]}\label{sec:CiderModel}

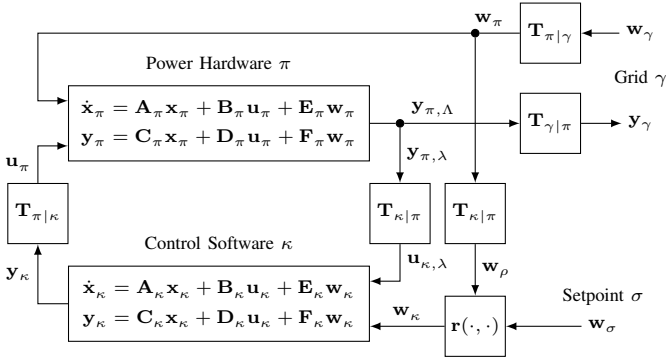
\begin{figure}[t]
    \centering
    {

\tikzstyle{system}=[rectangle, draw=black, minimum width=4.0cm, minimum height=1.0cm, inner sep=0pt]
\tikzstyle{block}=[rectangle, draw=black, minimum size=0.8cm, inner sep=0pt]
\tikzstyle{dot}=[circle, draw=black, fill=black, minimum size=0.1cm, inner sep=0pt]

\tikzstyle{signal}=[-latex]

\scriptsize

\begin{tikzpicture}

	\def\dx{1.0}
	\def\dy{1.0}
	
	
	
	\node[system] (P) at (0,1.2*\dy)
	{%
	$
	\begin{aligned}
		    \dot{\mathbf{x}}_{\pwr}
		&=      \mathbf{A}_{\pwr}\mathbf{x}_{\pwr}
		    +   \mathbf{B}_{\pwr}\mathbf{u}_{\pwr}
		    +   \mathbf{E}_{\pwr}\mathbf{w}_{\pwr}
	    \\
		    \mathbf{y}_{\pwr}
		&=      \mathbf{C}_{\pwr}\mathbf{x}_{\pwr}
		    +   \mathbf{D}_{\pwr}\mathbf{u}_{\pwr}
		    +   \mathbf{F}_{\pwr}\mathbf{w}_{\pwr}
	\end{aligned}
	$
	};
	
	\node at ($(P.north)+(0,0.3*\dy)$) {Power Hardware $\pwr$};
	
	\node[system] (C) at (0,-1.2*\dy)
	{%
	$
	\begin{aligned}
		    \dot{\mathbf{x}}_{\ctrl}
		&=      \mathbf{A}_{\ctrl}\mathbf{x}_{\ctrl}
		    +   \mathbf{B}_{\ctrl}\mathbf{u}_{\ctrl}
		    +   \mathbf{E}_{\ctrl}\mathbf{w}_{\ctrl}
	    \\
		    \mathbf{y}_{\ctrl}
		&=      \mathbf{C}_{\ctrl}\mathbf{x}_{\ctrl}
		    +   \mathbf{D}_{\ctrl}\mathbf{u}_{\ctrl}
		    +   \mathbf{F}_{\ctrl}\mathbf{w}_{\ctrl}
	\end{aligned}
	$
	};
	
	\node at ($(C.north)+(0,0.3*\dy)$) {Control Software $\ctrl$};
	
	\node[block] (TPC) at ($0.5*(P.east)+0.5*(C.east)+(0.4*\dx,0)$) {$\mathbf{T}_{\ctrl|\pwr}$};
	\node[block] (TCP) at ($0.5*(P.west)+0.5*(C.west)-(0.4*\dx,0)$) {$\mathbf{T}_{\pwr|\ctrl}$};
	
	\node[dot] (YP) at ($(P.east)+(0.4*\dx,0)$) {};
	
	\draw[-] (P.east) to (YP.west);
	\draw[signal] (YP.south) to node[right]{$\mathbf{y}_{\pwr,\lambda}$} (TPC.north);
	\draw[signal] (TPC.south)
		to node[right]{$\mathbf{u}_{\ctrl,\lambda}$} ($(C.east)+(0.4*\dx,0.3*\dy)$)
		to ($(C.east)+(0,0.3*\dy)$);
	
	\draw[signal] (C.west)
		to ($(C.west)-(0.4*\dx,0)$)
		to node[left]{$\mathbf{y}_{\ctrl}$} ($(TCP.south)$);
	\draw[signal] (TCP.north)
		to node[left]{$\mathbf{u}_{\pwr}$} ($(P.west)-(0.4*\dx,0.3*\dy)$)
		to ($(P.west)-(0,0.3*\dy)$);
		
	
	\node[block] (TPR) at ($(TPC)+(\dx,0)$) {$\mathbf{T}_{\ctrl|\pwr}$};
	\node[dot] (WP) at ($(TPR)+(0,2.4*\dy)$) {};
	\node[block] (TGP1) at ($(WP)+(\dx,0)$) {$\mathbf{T}_{\pwr|\grid}$};
	\node[block] (TGP2) at ($(TGP1)-(0,1.2*\dy)$) {$\mathbf{T}_{\grid|\pwr}$};
	\node (WG) at ($(TGP1)+(1.2*\dx,0)$) {$\mathbf{w}_{\grid}$};
	\node (YG) at ($(TGP2)+(1.2*\dx,0)$) {$\mathbf{y}_{\grid}$};
	
	\draw[signal] (WG.west) to (TGP1.east);
	\draw[signal] (TGP2.east) to (YG.west);
	\draw[-] (TGP1.west) to node[near end,above]{$\mathbf{w}_{\pwr}$} (WP.east) ;
	\draw[signal] (WP.south) to (TPR.north);
	\draw[signal] (WP.west)
		to ($(P.west)+(-0.4*\dx,1.2*\dy)$)
		to ($(P.west)+(-0.4*\dx,0.3*\dy)$)
		to ($(P.west)+(0,0.3*\dy)$);
	\draw[signal] (YP.east) to node[near start,above]{$\mathbf{y}_{\pwr,\Lambda}$} (TGP2.west);
	
	\node at ($0.5*(WG)+0.5*(YG)$) {Grid $\grid$};
	
	
	
	\node[block] (R) at ($(TPR)-(0,1.5*\dy)$) {$\mathbf{r}(\cdot,\cdot)$};
	\node (SP) at ($(R)+1.7*(\dx,0)$) {$\mathbf{w}_{\spt}$};
	
	\draw[signal] (SP.west) to (R.east);
	\draw[signal] (TPR.south) to node[right]{$\mathbf{w}_{\refr}$} (R.north);
	\draw[signal] (R.west) to node[above]{$\mathbf{w}_{\ctrl}$} ($(C.east)-(0,0.3*\dy)$);
	
	\node at ($(SP)+0.4*(0,\dy)$) {Setpoint $\spt$};
	
\end{tikzpicture}

}
    \caption
    {%
        Block diagram of the proposed generic state-space model of \CIDER[s].
        Note the modularity: power hardware $\pwr$, control software $\ctrl$, and grid $\grid$ are represented by separate blocks, which are interfaced via coordinate transformations.
        The reference calculation $\mathbf{r}(\cdot)$ may be either linear (i.e., for $\mathit{Vf}$ control) or nonlinear (i.e., for $\mathit{PQ}$ control).
        The other blocks of the model are exactly linear (i.e., \LTP systems and \LTP transforms). Adapted from~\cite{kettner2018properties}.}
    \label{fig:CIDER:model}
\end{figure}
In~\cite{kettner2021harmonic_part_I}, the authors developed in the time domain a generic \LTP model of the CIDERs regardless of their operative mode. CIDERs are classified as either  \textit{grid-forming} or \textit{grid-following}, according to whether they control the voltage's magnitude and frequency (i.e., $Vf$ control) or the injected current w.r.t. the fundamental component of the grid voltage at the point of connection (i.e., $PQ$ control).

As shown in~\cite{kettner2021harmonic_part_I}, a generic CIDER consists of \textit{the power hardware $\pi$} and \textit{the control software $\kappa$}, which are represented according to the following LTP model $\Sigma$:
\begin{equation}
    \Sigma : \left\{ \begin{array}{c}
         \dot{\mathbf{x}}(t) =   \mathbf{A}(t)\mathbf{x}(t) + \mathbf{B}(t)\mathbf{u}(t)+\mathbf{E}(t)\mathbf{w}(t)\\
          \mathbf{y}(t) =   \mathbf{C}(t)\mathbf{x}(t) + \mathbf{D}(t)\mathbf{u}(t)+\mathbf{F}(t)\mathbf{w}(t)
    \end{array}
    \right.
    \label{eq:CIDER:openloop}
\end{equation}
where $\mathbf{x}(t)$, $\mathbf{u}(t)$, $\mathbf{y}(t)$, $\mathbf{w}(t)$ are the \textit{state}, \textit{input}, \textit{output} and \textit{disturbance vector}, respectively. Moreover, $\mathbf{A}(t)$, $\mathbf{B}(t)$, $\mathbf{C}(t)$, $\mathbf{D}(t)$, $\mathbf{E}(t)$ and $\mathbf{F}(t)$ are the \textit{system}, \textit{output}, \textit{feed-through}, \textit{input disturbance} and \textit{output disturbance matrix}, respectively. As shown in \cref{fig:CIDER:model}, the generic \CIDER state-space model is obtained by combining the power hardware and control software models.

    
    It is important to note that $\pi$ and $\kappa$ may adopt two different reference frames (e.g., it is common practice to adopt the phase (ABC) coordinates for $\pi$ and the \textit{Direct-Quadrature} (DQ) coordinates for $\kappa$~\cite{rocabert2012control}). This requires the calculation of matrices to pass from one frame to another. Specifically, $\mathbf{T}_{\pi|\kappa}(t)$ and $\mathbf{T}_{\kappa|\pi}(t)$ perform the change of coordinates from the power hardware to the control software frame and vice-versa.  
    For instance, this happens in calculating the control software disturbance vector $\mathbf{w}_{k}(t)$, which is typically expressed as a function of the power hardware disturbance vector $\mathbf{w
    }_{\pi}(t)$ and the \CIDER setpoint vector as follows:
\begin{equation}
    \mathbf{w}_{\kappa}(t) = \mathbf{r}\left(\mathbf{T}_{\kappa|\pi}(t)\mathbf{w}_{\pi}(t),\mathbf{w}_{\sigma}(t)\right)
\end{equation}
where 
\begin{align}
    \mathbf{w}_{\spt}(t)
	&\sim	\left\{
				\begin{array}{cl}
					V,f &\text{if \CIDER is grid-forming}	\\
					P,Q &\text{if \CIDER is grid-following}
				\end{array}
				\right.
	\label{eq:setpoint:time}
\end{align}
Additionally, it is worth noting that the function $\mathbf{r}(\cdot)$ may be non-linear. More precisely, grid-forming \CIDER[s] (i.e., with $Vf$ control) are characterized by a linear function \(\textbf{r}(\cdot)\), whereas in the case of grid-following \CIDER[s] (i.e., with $PQ$ control) the function $\mathbf{r}(\cdot)$ is non-linear.

    Depending on the type of the \CIDER connection with the grid (e.g., the grid is usually a four-wire system whereas the \CIDER[s] can be either three-leg or four-leg devices), it is necessary to perform the change of coordinates from the grid to the power hardware and vice-versa (i.e., $\mathbf{T}_{\gamma|\pi}(t)$ and $\mathbf{T}_{\pi|\gamma}(t)$). This is particularly relevant in the calculation of $\mathbf{w}_{\pi}(t)$, which is given by the grid disturbance vector \(\mathbf{w}_{\gamma}(t)\):
 \begin{equation}
    \mathbf{w}_{\pi}(t) = \mathbf{T}_{\pi|\gamma}(t)\mathbf{w}_{\gamma}(t)
\end{equation}
It is worth noting that, depending on the nature of the \CIDER, the grid disturbance and output vectors represent either the nodal injected current or the nodal voltage, as summarised below:
\begin{align}
    \mathbf{w}_{\grid}(t)
	&\sim	\left\{
				\begin{array}{cl}
					\mathbf{i}(t)	&\text{if \CIDER is grid-forming}	\\
					\mathbf{v}(t)	&\text{if \CIDER is grid-following}
				\end{array}
				\right.
	\label{eq:grid:disturbance:time}\\
	\mathbf{y}_{\grid}(t)
	&\sim	\left\{
				\begin{array}{cl}
					\mathbf{v}(t)	&\text{if \CIDER is grid-forming}	\\
					\mathbf{i}(t)	&\text{if \CIDER is grid-following}
				\end{array}
				\right.
	\label{eq:grid:output:time}
\end{align}

As shown in Fig.~\ref{fig:CIDER:model}, the open-loop system $\Sigma$ form a closed-loop system via the feedback matrix $\mathbf{T}(t)$:
\begin{equation}
    \mathbf{u}(t) = \mathbf{T}(t)\mathbf{y}(t)\label{eq:CIDER:feedback}
\end{equation}
More precisely, $\mathbf{T}(t)$ is a block matrix, whose blocks are  $\mathbf{T}_{\pi|\kappa}(t)$ and $\mathbf{T}_{\kappa|\pi}(t)$.

The time-periodic nature of \eqref{eq:CIDER:openloop} suggests employing the Fourier series for the expansion of vectors and matrices. It is important to note that the product of waveforms in the time domain results in the convolution of their spectra in the frequency domain. In particular, the model above involves the product between the time-varying matrices and vectors. In the frequency domain, this results in the product between Toeplitz matrices and vectors of Fourier coefficients.

The \CIDER model in the harmonic domain involves calculating its internal response by combining~\eqref{eq:CIDER:openloop} with~\eqref{eq:CIDER:feedback} when expressed in the frequency domain.
\begin{definition}\label{def:CIDER:int_resp}
    The internal response of the CIDER describes the link from $\mathbf{w}(t)$ to $\mathbf{y}(t)$.
\end{definition}
It is given by the following expression in the frequency domain:
\begin{equation}\label{eq:CIDER:output}
    \hat{\Y}(\hat{\W}) = \hat{\G}\hat{\W}
\end{equation}
where  $\hat{\Y}$ and $\hat{\W}$ are the vectors of the  Fourier coefficients of $\mathbf{y}(t)$ and $\mathbf{w}(t)$, respectively, whereas $\hat{\G}$ is the closed-loop gain matrix of the \CIDER in the harmonic domain. \eqref{eq:CIDER:output} can be conveniently rearranged in the following block form:
\begin{equation}
    \begin{bmatrix}
        \hat{\Y}_{\pi} \\ \hat{\Y}_{\kappa}
    \end{bmatrix} = \begin{bmatrix}
                \Hat{\mathbf{G}}_{\pwr\pwr}
            &   \Hat{\mathbf{G}}_{\pwr\ctrl}\\
                \Hat{\mathbf{G}}_{\ctrl\pwr}
            &   \Hat{\mathbf{G}}_{\ctrl\ctrl}  
\end{bmatrix}
\begin{bmatrix}
        \hat{\W}_{\pi} \\ \hat{\W}_{\kappa}
    \end{bmatrix}
\label{eq:CIDER:transfer:block-form}
\end{equation}
where $\hat{\W}_{\pi}$ and $\hat{\W}_{\kappa}$ are the Fourier coefficients of $\mathbf{w}_{\pi}$ and $\mathbf{w}_{\kappa}$, respectively. Remarkably, in the case of grid-forming \CIDER[s] the system of equations in \eqref{eq:CIDER:transfer:block-form} is linear, whereas for grid-following \CIDER[s] it is partly non-linear in the variable $\hat{\W}_{\kappa}$.

Finally, in order to implement the \HPF, the computation of the \CIDER \textit{grid response} is required.  
\begin{definition}\label{def:CIDER:grid_resp}
 The grid response of the \CIDER represents the the link from $\mathbf{w}_{\gamma}(t)$ and $\mathbf{w}_{\sigma}(t)$ to $\mathbf{y}_{\gamma}(t)$.
\end{definition} 

As shown in~\cite{kettner2021harmonic_part_I}, the grid response evaluation involves the calculation of several quantities, namely the \CIDER internal response, the reference $\mathbf{r}(\cdot)$ and the grid-side transformations $\mathbf{T}_{\gamma,\pi}$ and $\mathbf{T}_{\pi,\gamma}$, as follows:
\begin{equation}\label{eq:CIDER:transfer:outer}
    \Hat{\mathbf{Y}}_{\gamma}(\Hat{\mathbf{W}}_{\gamma}, \Hat{\mathbf{W}}_{\sigma}) = \Hat{\mathbf{T}}_{\gamma|\pi}\Hat{\mathbf{Y}}_{\pi}(\Hat{\mathbf{T}}_{\pi|\gamma}\Hat{\mathbf{W}}_{\gamma},\Hat{\mathbf{W}}_{\sigma}) 
\end{equation}
where $\Hat{\mathbf{T}}_{\gamma|\pi}$ and $\Hat{\mathbf{T}}_{\pi|\gamma}$ are the Toeplitz matrices composed of the Fourier coefficient of $\mathbf{T}_{\gamma|\pi}$ and $\mathbf{T}_{\pi|\gamma}$, respectively. Moreover, $\Hat{\mathbf{Y}}_{\pi}$ can be easily deduced from~\eqref{eq:CIDER:transfer:block-form}. More precisely, the \CIDER grid response \eqref{eq:CIDER:transfer:outer}, depending on the nature of the resource, results in the following expressions (grid-forming and grid-following, respectively):
\begin{alignat}{4}
    s\in \formers:~& \Hat{\mathbf{Y}}_{\gamma,s}\coloneqq~\Hat{\V}_s&~=~&\bar{\mathbf{G}}_{\pwr\pwr,s}\Hat{\I}_{s}+\bar{\mathbf{G}}_{\pwr\ctrl,s}\hat{\W}_{\kappa,s}\label{eq:CIDER:formingclosedloop}\\
     r\in \followers:~&\Hat{\mathbf{Y}}_{\gamma,r}\coloneqq~\Hat{\I}_r&~=~&\bar{\mathbf{G}}_{\pwr\pwr,r}\Hat{\V}_{r}+\bar{\mathbf{G}}_{\pwr\ctrl,r}\hat{\W}_{\kappa,r}\label{eq:CIDER:followingclosedloop}
\end{alignat}
where $\forall~n\in \formers \cup \followers$
\begin{alignat}{1}
  \bar{\mathbf{G}}_{\pwr\pwr} &=  \Hat{\mathbf{T}}_{\gamma|\pi}\Hat{\mathbf{G}}_{\pwr\pwr}\mathbf{\Hat{T}}_{\pi|\gamma}\\
   \bar{\mathbf{G}}_{\pwr\ctrl} &= \Hat{\mathbf{T}}_{\gamma|\pi}\Hat{\mathbf{G}}_{\pwr\ctrl}
\end{alignat}

\subsection{Mathematical Formulation of the \HPF Problem}

The \HPF framework proposed in~\cite{kettner2021harmonic_part_I} is based on the mismatch equations, which must be equal to zero in equilibrium at the Point of Common Coupling (PCC) between the network and \CIDER[s]. More precisely, they involve nodal equations from both the grid's and the \CIDER[s'] perspective. The unknown variables are the nodal injected currents $\Hat{\mathbf{I}}_{\formers}$ at nodes $\formers$ and the nodal phase-to-ground voltages $\Hat{\mathbf{V}}_{\followers}$  at $\followers$.

From the point of view of the grid, the nodal equations are given by the hybrid parameters defined in~\eqref{eq:nodes:hybrid:harmonics}:
\begin{align}
        \Hat{\mathbf{V}}_{\formers}(\Hat{\mathbf{I}}_{\formers},\Hat{\mathbf{V}}_{\followers})
    &=      \Hat{\mathbf{H}}_{\formers\times\formers}\Hat{\mathbf{I}}_{\formers}
        +   \Hat{\mathbf{H}}_{\formers\times\followers}\Hat{\mathbf{V}}_{\followers}
    \label{eq:grid:form}
    \\
        \Hat{\mathbf{I}}_{\followers}(\Hat{\mathbf{I}}_{\formers},\Hat{\mathbf{V}}_{\followers})
    &=      \Hat{\mathbf{H}}_{\followers\times\formers}\Hat{\mathbf{I}}_{\formers}
        +   \Hat{\mathbf{H}}_{\followers\times\followers}\Hat{\mathbf{V}}_{\followers}
    \label{eq:grid:follow}
\end{align}
From the point of view of the \CIDER[s], the nodal equations are established via the closed-loop transfer function~\eqref{eq:CIDER:transfer:outer} as follows:
\begin{alignat}{2}
        s\in\formers
    &:  ~
    &   \Hat{\mathbf{V}}_{s}(\Hat{\mathbf{I}}_{s},V_{\spt,s},f_{\spt,s})
    &=  \Hat{\mathbf{Y}}_{\grid,s}(\Hat{\mathbf{T}}_{\pwr|\grid}\Hat{\mathbf{I}}_{s},V_{\spt,s},f_{\spt,s})
    \label{eq:resource:form}\\
        r\in\followers
    &:  ~
    &   \Hat{\mathbf{I}}_{r}(\Hat{\mathbf{V}}_{r},S_{\spt,r})
    &=  \Hat{\mathbf{Y}}_{\grid,r}(\Hat{\mathbf{T}}_{\pwr|\grid}\Hat{\mathbf{V}}_{r},S_{\spt,r})
    \label{eq:resource:follow}
\end{alignat}

The mismatches between \eqref{eq:grid:form}--\eqref{eq:grid:follow} and \eqref{eq:resource:form}--\eqref{eq:resource:follow} must be zero in equilibrium:
\begin{align}
        \Delta\Hat{\mathbf{V}}_{\formers}
        (\Hat{\mathbf{I}}_{\formers},\Hat{\mathbf{V}}_{\followers},\mathbf{V}_{\spt},\mathbf{f}_{\spt})
    &=  \mathbf{0}
    \label{eq:residual:form}\\
        \Delta\Hat{\mathbf{I}}_{\followers}
        (\Hat{\mathbf{I}}_{\formers},\Hat{\mathbf{V}}_{\followers},\mathbf{S}_{\spt})
    &=  \mathbf{0}
    \label{eq:residual:follow}
\end{align}
where $\mathbf{V}_{\spt}$, $\mathbf{f}_\spt$, and $\mathbf{S}_{\spt}$ are column vectors built of $V_{\spt,s}$, $f_{\spt,s}$ ($s\in\formers$) and $S_{\spt,r}$ ($r\in\followers$), respectively.

It is worth noting that~\eqref{eq:residual:form} and~\eqref{eq:residual:follow} constitute a non-linear system of equations. In~\cite{kettner2021harmonic_part_I}, the authors presented a numerical algorithm based on the \NR method to find the solution to this problem.

\section{Fixed-Point Formulation and Uniqueness of the Solution}\label{sec:Fixed-PointFormulation}
This section aims to derive the fixed-point formulation of the \HPF problem, in the form:
\begin{equation}
    \mathbf{x} = {\boldsymbol \Phi}(\mathbf{x})
\end{equation}
where $\mathbf{x}$ is the vector of unknown variables and $\boldsymbol \Phi$ is the function accounting for the mismatch equations. This formulation will be used in the following to prove the uniqueness of the solution based on the contraction property of $\boldsymbol \Phi$.

In order to express the \HPF as a fixed-point problem, the first step consists of appropriately sorting the vectors $\hat{\I}_{\formers}$ and $\hat{\V}_{\followers}$. From the grid’s perspective, it seems natural to sort generic quantities, as described by \eqref{eq:Hybrid:Voltage}, first based on their nature (e.g., grid-forming or grid-following) and, then, according to their harmonic order.  More precisely, let  \(\mathbf{X}_{o}\) represent a generic quantity (e.g., nodal current or voltage) where \(o\in \mathcal{O}\)  denotes the index of the subset to which it belongs (e.,g, $\formers$ or $\followers$). This sorting operation can be expressed as follows:
\begin{equation}\label{eq:sorting1}
    \hat{\mathbf{X}}_{\mathcal{O}} = \col_{h\in \harmonics}\col_{o\in \mathcal{O}}(\mathbf{X}_{o}(f_h))
\end{equation}
Conversely, from the point of view of \CIDER[s], it seems more intuitive to sort quantities based on their harmonic order first and, then, according to their nature, namely:
\begin{equation}\label{eq:sorting2}
\Tilde{\mathbf{X}}_{\mathcal{O}}= \col_{o\in \mathcal{O}}\col_{h\in \harmonics}({\mathbf{X}}_{o}(f_h))
\end{equation}
Hence, to establish an overall coherent model, it is necessary to choose between \eqref{eq:sorting1} and \eqref{eq:sorting2} sorting criteria. In the following the second criterion has been adopted. This results in
\begin{alignat}{2}
    \Tilde{\I}_{\formers} &\coloneqq          \col_{s\in\formers}(\hat{\I}_s) &~\in~   \CompNum^{3\Card{\harmonics}\Card{\formers}\times1}\\
    \Tilde{\V}_{\followers} &\coloneqq          \col_{r\in\followers}(\hat{\V}_r) &~\in~   \CompNum^{3\Card{\harmonics}\Card{\followers}\times1}
\end{alignat}
The sorting scheme can be expressed compactly as:
\begin{alignat}{2}
    \Tilde{\I}_{\formers} & = \PermutationFormer\hat{\I}_{\formers} \\  
    \Tilde{\V}_{\followers}& = \PermutationFollower\hat{\V}_{\followers}
\end{alignat}
where $\PermutationFormer \in \left\{0,1\right\}^{3\Card{\harmonics}\Card{\formers}\times3\Card{\harmonics}\Card{\formers}}$ and $\PermutationFollower \in \left\{0,1\right\}^{3\Card{\harmonics}\Card{\followers}\times3\Card{\harmonics}\Card{\followers}}$ are the permutation matrices obtained by
permuting the columns of the identity matrix. More precisely, the permutation matrix $\mathbf{P}_{\formers}$ is block-wise defined as follows~\cite{becker2024unified}:
\begin{equation}
    \mathbf{P}_{\formers, ij} = \left\{
				\begin{array}{cl}
					\diag{\left(\mathbf{1}_{3}\right)} & \forall s\in \formers, \forall h\in \harmonics:  	\\
     & i = (s-1)\Card{\harmonics} + h + (h_{\text{max}}+1),\\
     & j = (h+h_{\text{max}})\Card{\formers} + s\\ 
					\mathbf{0}_{3\times 3} & \text{otherwise}
				\end{array}
				\right.
\end{equation} 
where \(\diag{\left(\mathbf{1}_{3}\right)}\) and \(\mathbf{0}_{3\times 3}\) are the identity and null matrices of size \(3 \times 3\), respectively, whereas \(h_{\text{max}}\) denotes the highest harmonic component of the frequency span considered in the \HPF study.   The matrix $\mathbf{P}_{\followers}$ is defined analogously. Hence, the overall sorting matrix is defined as follows
\begin{equation}
    \mathbf{P} = \diag{(\PermutationFormer,\PermutationFollower)}
\end{equation}
Remarkably, the adopted sorting scheme induces permutations within the rows and the columns of $\hat{\mathbf{H}}$, resulting in\footnote{The inverse of a permutation matrix exists and is given by its transpose.}
\begin{equation}
    \Tilde{\mathbf{H}} = \mathbf{P}\hat{\mathbf{H}}\mathbf{P}^T
\end{equation}
This leads \eqref{eq:nodes:hybrid:harmonics} to be restated as follows:
\begin{equation}
    \begin{bmatrix}
	        \Tilde{\V}_{\formers}\\
	        \Tilde{\I}_{\followers}
	    \end{bmatrix}
    =   \begin{bmatrix}
	            \Tilde{\HG}_{\formers\times\formers}
            &   \Tilde{\HG}_{\formers\times\followers}\\
                \Tilde{\HG}_{\followers\times\formers}
            &   \Tilde{\HG}_{\followers\times\followers}
	    \end{bmatrix}
	    \begin{bmatrix}
	        \Tilde{\I}_{\formers}\\
	        \Tilde{\V}_{\followers}
	    \end{bmatrix}
	   \label{eq:nodes:hybrid:harmonicsnew}
\end{equation}
where the matrix's blocks are defined as follows:
\begin{alignat}{2}
    \Tilde{\HG}_{\formers\times\formers} &= \PermutationFormer \hat{\HG}_{\formers\times\formers}\PermutationFormer^T\\
    \Tilde{\HG}_{\formers\times\followers} &= \PermutationFormer \hat{\HG}_{\formers\times\followers}\PermutationFollower^T\\
    \Tilde{\HG}_{\followers\times\formers} & = \PermutationFormer\hat{\HG}_{\followers\times\formers}\PermutationFollower^T\\
    \Tilde{\HG}_{\followers\times\followers} &= \PermutationFollower\hat{\HG}_{\followers\times\formers}\PermutationFollower^T
\end{alignat}
Thus, the mismatch equations \eqref{eq:residual:form} and \eqref{eq:residual:follow} can be restated as follows
\begin{alignat}{1}
\Tilde{\HG}_{\formers\times\formers}\Tilde{\I}_{\formers}+\Tilde{\HG}_{\formers\times\followers}\Tilde{\V}_{\followers}-\Tilde{\mathbf{G}}_{\pwr\pwr,\formers}\Tilde{\I}_{\formers}-\Tilde{\mathbf{G}}_{\pwr\ctrl,\formers}\Tilde{\W}_{\kappa,\formers} = \mathbf{0}\label{eq:residual:formnew}\\
\Tilde{\HG}_{\followers\times\formers}\Tilde{\I}_{\formers}+\Tilde{\HG}_{\followers\times\followers}\Tilde{\V}_{\followers}-\Tilde{\mathbf{G}}_{\pwr\pwr,\followers}\Tilde{\V}_{\followers}-\Tilde{\mathbf{G}}_{\pwr\ctrl,\followers}\Tilde{\W}_{\kappa,\followers} = \mathbf{0}\label{eq:residual:follownew}
\end{alignat}
where the following can be inferred from \eqref{eq:CIDER:formingclosedloop} and \eqref{eq:CIDER:followingclosedloop}:
\begin{alignat}{2}
    \Tilde{\mathbf{G}}_{\pwr\pwr,\formers}&\coloneqq \diag_{s\in\formers}(\bar{\mathbf{G}}_{\pwr\pwr,s})&~\Tilde{\mathbf{G}}_{\pwr\pwr,\followers}&\coloneqq \diag_{r\in\followers}(\bar{\mathbf{G}}_{\pwr\pwr,r})\\
    \Tilde{\mathbf{G}}_{\pwr\ctrl,\formers}&\coloneqq \diag_{s\in\formers}(\bar{\mathbf{G}}_{\pwr\ctrl,s})&~
    \Tilde{\mathbf{G}}_{\pwr\ctrl,\followers}&\coloneqq \diag_{r\in\followers}(\bar{\mathbf{G}}_{\pwr\ctrl,r})\\
    \Tilde{\W}_{\kappa,\formers}&\coloneqq \col_{s\in\formers}(\bar{\W}_{\kappa,s})&~
    \Tilde{\W}_{\kappa,\followers}&\coloneqq \col_{r\in\followers}(\bar{\W}_{\kappa,r})
\end{alignat}
\begin{remark}
In a grid-following converter, the current injected by the converter is controlled with a specific phase displacement w.r.t. the grid voltage at the PCC. Therefore, the control software disturbance can be expressed as a function of the nodal voltage and the reference power $\mathbf{S}_{\sigma,\followers}$:
\begin{equation}\label{eq:ControlDisturbanceVoltage}
  \Tilde{\W}_{\kappa,\followers} = \Tilde{\W}_{\kappa,\followers}(\Tilde{\mathbf{T}}_{\ctrl|\pwr}\Tilde{\mathbf{T}}_{\pwr|\gamma}\hat{\mathbf{V}}_\followers,\mathbf{S}_{\sigma,\followers})
\end{equation}   
The grid-forming unit regulates the voltage magnitude and angle at the PCC and controls the network frequency. As a result, the disturbance in the control software depends on the reference voltage $\mathbf{V}_{\sigma,\formers}$ and frequency:
\begin{equation}
    \Tilde{\W}_{\kappa,\formers} = \Tilde{\W}_{\kappa,\formers}(\mathbf{V}_{\sigma,\formers},f_{\sigma})
    \end{equation}
\end{remark}

The system of equations involving \eqref{eq:residual:formnew} and \eqref{eq:residual:follownew} features  $\Tilde{\mathbf{I}}_{\formers}$ and $\Tilde{\mathbf{V}}_{\followers}$ as unknowns variables. However, it can be reformulated to get a reduced fixed-point model in the single variable
$\Tilde{\mathbf{V}}_{\followers}$. This can be achieved, by inverting the relationship in \eqref{eq:residual:formnew}\footnote{Similar reasoning holds for the formulation in the variable $\Tilde{\mathbf{I}}_{\formers}$, which would require inverting the relationship in \eqref{eq:residual:follownew}. However, this operation may not be straightforward due to the dependence of $\Tilde{\W}_{\ctrl,\followers}$ on $\Tilde{\mathbf{V}}_\followers$.}.
In this regard, the following two conditions must be satisfied:
\begin{condition}\label{cond:1} 
The matrix $\Tilde{\mathbf{L}}_{\formers\times \formers}$, defined as follows, is invertible
        \begin{equation}    \Tilde{\mathbf{L}}_{\formers\times \formers} = \Tilde{\mathbf{G}}_{\pwr\pwr,\formers}-\Tilde{\HG}_{\formers\times\formers}
    \end{equation}
\end{condition}
\begin{condition} \label{cond:2} The matrix $\Tilde{\mathbf{K}}_{\followers\times\followers}$, defined as follows, is invertible
    \begin{equation}
    \Tilde{\mathbf{K}}_{\followers\times\followers} = \Tilde{\mathbf{H}}_{\followers\times \followers}-\Tilde{\mathbf{G}}_{\pwr\pwr,\followers}+\Tilde{\mathbf{H}}_{\followers\times\formers}\Tilde{\mathbf{L}}_{\formers\times \formers}^{-1}\Tilde{\mathbf{H}}_{\formers\times \followers}\end{equation}
\end{condition}
It is important to note that Condition~\ref{cond:1} and Condition~\ref{cond:2} are fundamental requirements for the adoption of a fixed-point-based algorithm. In particular, the matrices $\Tilde{\mathbf{L}}_{\formers\times \formers}$ and $\Tilde{\mathbf{K}}_{\followers\times\followers}$ are dependent on the structural properties of the grid and the selected control parameters. Therefore, their invertibility needs to be assessed only once, ex-ante, (i.e., before applying the fixed-point algorithm). Hence, once the Condition~\ref{cond:1} has been verified, the $\Tilde{\I}_{\formers}$ can be expressed as follows:  
\begin{equation}
    \Tilde{\I}_{\formers}= \Tilde{\mathbf{L}}_{\formers\times \formers}^{-1}\left(\Tilde{\HG}_{\formers\times\followers}\Tilde{\V}_{\followers}-\Tilde{\mathbf{G}}_{\pwr\ctrl,\formers}\Tilde{\W}_{\kappa,\formers}\right) \label{eq:residual:currentformnew}\end{equation}
By substituting \eqref{eq:residual:currentformnew} into \eqref{eq:residual:follownew} and assuming that Condition~\ref{cond:2} holds, the fixed-point formulation of the \HPF in the variable $\Tilde{\mathbf{V}}_\followers$ is obtained: 
\begin{equation}\label{eq:HPF:FixedPoint}
\Tilde{\mathbf{V}}_\followers = \Tilde{\mathbf{K}}_{\followers\times\followers}^{-1}\left(\Tilde{\mathbf{G}}_{\pwr\ctrl,\followers}\Tilde{\W}_{\kappa,\followers}-\Tilde{\mathbf{K}}_{\followers\times\formers}\Tilde{\mathbf{G}}_{\pwr\ctrl,\formers}\Tilde{\W}_{\kappa,\formers}\right)
\end{equation}
where 
  \begin{equation}
        \Tilde{\mathbf{K}}_{\followers\times\formers} = \Tilde{\mathbf{H}}_{\followers\times\formers}\Tilde{\mathbf{L}}_{\formers\times \formers}^{-1}
    \end{equation}
 Note that the fixed-point nature of \eqref{eq:HPF:FixedPoint} lies in the dependence of $\Tilde{\W}_{\ctrl,\followers}$ on $\Tilde{\mathbf{V}}_\followers$. More precisely, as highlighted by \eqref{eq:ControlDisturbanceVoltage}, the control software disturbance $\W_{\ctrl,\followers}$ depends on the quantity $\Tilde{\mathbf{T}}_{\ctrl|\pwr}\Tilde{\mathbf{T}}_{\pwr|\gamma}\hat{\mathbf{V}}_\followers$ that, according to the scheme in Fig.~\ref{fig:CIDER:model}, is denoted with $\Tilde{\W}_{\rho}$. Hence, it is more convenient to restate the fixed-point model in the variable $\Tilde{\W}_{\rho}$ as follows:
 \begin{equation}
     \Tilde{\W}_{\rho} = \Tilde{\mathbf{B}}_{\followers\times\followers}\Tilde{\W}_{\kappa,\followers}(\Tilde{\W}_{\rho},\mathbf{S}_{\sigma,\followers})-\Tilde{\mathbf{B}}_{\followers\times\formers}\Tilde{{\W}}_{\kappa,\formers}(\mathbf{V}_{\sigma,\formers},f_\sigma)
 \end{equation}
 where 
 \begin{equation}
     \Tilde{\mathbf{B}}_{\followers\times\followers} = \Tilde{\mathbf{T}}_{\ctrl|\gamma}\Tilde{\mathbf{K}}_{\followers\times\followers}^{-1}\Tilde{\mathbf{G}}_{\pwr\ctrl,\followers}
 \end{equation}
  \begin{equation}
     \Tilde{\mathbf{B}}_{\followers\times\formers} = \Tilde{\mathbf{T}}_{\ctrl|\gamma}\Tilde{\mathbf{K}}_{\followers\times\followers}^{-1}\Tilde{\mathbf{K}}_{\followers\times\formers}\Tilde{\mathbf{G}}_{\pwr\ctrl,\formers}
 \end{equation}
Finally, the \HPF solution is found through the fixed-point iteration:
\begin{equation}
    \Tilde{\W}_{\rho}^{(k+1)} = \boldsymbol{\Phi}(\Tilde{\W}_{\rho}^{(k)})
\end{equation}
The iterations stop when both the errors on   \(\Tilde{\W}_{\rho}^{(k+1)}\) and \(\boldsymbol{\Phi}\left(\Tilde{\W}_{\rho}^{(k+1)}\right)\) are less than given tolerances (e.g., \(\delta_x>0\) and \(\delta_f>0\) ). In other words, the following  conditions must be met:
\begin{equation}\label{eq:convergence}
    \begin{cases}
    \| \Tilde{\W}_{\rho}^{(k+1)}- \Tilde{\W}_{\rho}^{(k)}\|_\infty \leq \delta_x^* \\
    \| \boldsymbol{\Phi}(\Tilde{\W}_{\rho}^{(k)})\|_\infty \leq \delta_f^*
\end{cases}
\end{equation}
where \(\|\cdot\|_{\infty}\) denotes the infinity norm.

Once the convergence criteria in \eqref{eq:convergence} are met, the uniqueness of the solution can be assessed via the following theorem: 
\begin{Theorem}\label{Th:1}
Assume that \(\boldsymbol{\Phi}\) is continuously differentiable in a neighbourhood of a fixed point \(\Tilde{\W}^*_{\rho}\) of \(\boldsymbol{\Phi}\) and $\|\nabla\boldsymbol{\Phi}(\Tilde{\W}^*_{\rho})\|_{\infty}~<~1$. Then there exists a closed neighborhood \(\Omega\) of \(\Tilde{\W}_{\rho}\) such that \(\boldsymbol{\Phi}\) is a contraction on \(\Omega\). In particular, the fixed point iteration \(\Tilde{\W}_{\rho}^{(k+1)}=\boldsymbol{\Phi}(\Tilde{\W}_{\rho}^{(k)})\) converges for every \(\Tilde{\W}_{\rho}^{(0)}\in \Omega\) to \(\Tilde{\W}^*_{\rho}\)
\end{Theorem}
This theorem is a straightforward consequence of the Banach-Caccioppoli contraction theorem~\cite{quarteroni2006numerical}. 

Thus, the assessment of the uniqueness of the solution relies on computing the  Jacobian of \(\boldsymbol{\Phi}\) w.r.t. \(\Tilde{\W}_{\rho}\):
\begin{equation}\label{eq:Jacobian}
    \nabla \boldsymbol{\Phi} = \Tilde{\mathbf{B}}_{\followers\times\followers}\frac{\partial}{\partial \Tilde{\W}_{\rho}}\Tilde{\W}_{\kappa,\followers}(
    \Tilde{\W}_{\rho})
\end{equation}
and proving that the infinity norm of the quantity in \eqref{eq:Jacobian} is smaller than one for \(\Tilde{\W}_{\rho} = \Tilde{\W}^*_{\rho}\).

\section{Numerical Validation}\label{sec:NumericalValidation}
\begin{figure}[]
	\centering
    {
\tikzstyle{bus}=[circle,fill=black,minimum size=1.5mm,inner sep=0mm]
\tikzstyle{UG1}=[]
\tikzstyle{UG3}=[dashed]
\tikzstyle{load}=[-latex]

\begin{circuitikz}
	\scriptsize
	
	\def\BlockSize{1.0}	
	\def\X{3.8}	
	\def\Y{3.6}	
	\def\dlX{0.77}
	\def\dlY{1.1}
	\def\dload{0.4}
	
	
	\node[bus,label={left:N1}] (N1) at (0,0) {};
	\node[bus,label={below right:N2}] (N2) at (\dlX,0) {};
	\node[bus,gray,label={below:N21}] (N21) at (\dlX,-\dlY) {};
	\node[bus,label={below:N3}] (N3) at (2*\dlX,0) {};
	\node[bus,label={above:N11}] (N11) at (2*\dlX,\dlY) {};
	\node[bus,label={below:N4}] (N4) at (3*\dlX,0) {};
	\node[bus,label={below right:N5}] (N5) at (4*\dlX,0) {};
	\node[bus,label={left:N12}] (N12) at (4*\dlX,-\dlY) {};
	\node[bus,gray,label={below:N19}] (N19) at (5*\dlX,-\dlY) {};
	\node[bus,label={below:N13}] (N13) at (4*\dlX,-2*\dlY) {};
	\node[bus,gray,label={below:N22}] (N22) at (3*\dlX,-2*\dlY) {};
	\node[bus,label={below:N14}] (N14) at (5*\dlX,-2*\dlY) {};
	\node[bus,label={below:N15}] (N15) at (6*\dlX,-2*\dlY) {};
	\node[bus,label={below:N6}] (N6) at (5*\dlX,0) {};
	\node[bus,label={above:N16}] (N16) at (5*\dlX,\dlY) {};
	\node[bus,label={below:N7}] (N7) at (6*\dlX,0) {};
	\node[bus,label={below:N8}] (N8) at (7*\dlX,0) {};
	\node[bus,gray,label={above:N20}] (N20) at (7*\dlX,\dlY) {};
	\node[bus,label={below right:N9}] (N9) at (8*\dlX,0) {};
	\node[bus,label={below:N10}] (N10) at (8*\dlX,-\dlY) {};
	\node[bus,label={below:N17}] (N17) at (9*\dlX,0) {};
	\node[bus,label={below:N18}] (N18) at (9*\dlX,-\dlY) {};

	
	\draw[UG1] (N1) to node[midway,above]{35m} (N2) {};
	\draw[UG1] (N2) to node[midway,above]{35m} (N3) {};
	\draw[UG1] (N3) to node[midway,above]{35m} (N4) {};
	\draw[UG1] (N4) to node[midway,above]{35m} (N5) {};
	\draw[UG1] (N5) to node[midway,above]{35m} (N6) {};
	\draw[UG1] (N6) to node[midway,above]{35m} (N7) {};
	\draw[UG1] (N7) to node[midway,above]{35m} (N8) {};
	\draw[UG1] (N8) to node[midway,above]{35m} (N9) {};
	\draw[UG1] (N10) to node[sloped,anchor=center,above]{35m} (N9) {};
	
	\draw[UG3] (N3) to node[sloped,anchor=center,above]{30m} (N11) {};
	
	\draw[UG3] (N12) to node[sloped,anchor=center,above]{35m} (N5) {};
	\draw[UG3] (N13) to node[sloped,anchor=center,above]{35m} (N12) {};
	\draw[UG3] (N13) to node[midway,above]{35m} (N14) {};
	\draw[UG3] (N14) to node[midway,above]{30m} (N15) {};
	
	\draw[UG3] (N6) to node[sloped,anchor=center,above]{30m} (N16) {};
	\draw[UG3] (N9) to node[sloped,anchor=center,above]{30m} (N17) {};
	\draw[UG3] (N18) to node[sloped,anchor=center,above]{30m} (N10) {};

	\filldraw[gray,UG3] (N12) to node[sloped,anchor=center,above]{30m} (N19) {};
	\draw[gray,UG3] (N8) to node[sloped,anchor=center,above]{30m} (N20) {};
	\draw[gray,UG3] (N21) to node[sloped,anchor=center,above]{30m} (N2) {};
	\draw[gray,UG3] (N22) to node[midway,above]{30m} (N13) {};
	
	\draw[load] (N11) to node[right,align=left]{~P/Q}
	($(N11)+\dload*(1,0)$);
	\draw[load] (N15) to node[right,align=left]{~P/Q}
	($(N15)+\dload*(1,0)$);
	\draw[load] (N16) to node[right,align=left]{~P/Q}
	($(N16)+\dload*(1,0)$);
	\draw[load] (N17) to node[right,align=left]{~P/Q}
	($(N17)+\dload*(1,0)$);
	\draw[load] (N18) to node[right,align=left]{~V=230V\\~f=50Hz}
	($(N18)+\dload*(1,0)$);
	
	\draw[load] (N19) to node[right,align=left]{~Z}
	($(N19)+\dload*(1,0)$);
	\draw[load] (N20) to node[right,align=left]{~Z}
	($(N20)+\dload*(1,0)$);
	\draw[load] (N21) to node[right,align=left]{~Z}
	($(N21)+\dload*(1,0)$);
	\draw[load] (N22) to node[left,align=right]{Z~}
    ($(N22)-\dload*(1,0)$);

	\draw[-] ($(N1)+\dload*(0,-1)$) to ($(N1)+\dload*(0,1)$);
	\node[label={above:Substation}] (Substation) at ($(N1)+\dload*(0,1)$) {};

	
    \coordinate (Leg) at (-0.3*\dlX,-1.5*\dlY);
    \matrix [draw,below right] at (Leg) {
        \node [UG1,label=right:~~~~UG1] {}; \\
        \node [UG3,label=right:~~~~UG3] {}; \\
    };
    \draw[UG1] ($(Leg)+(0.15*\dlX,-0.25*\dlY)$) to node[]{} ($(Leg)+(0.75*\dlX,-0.25*\dlY)$);
    \draw[UG3] ($(Leg)+(0.15*\dlX,-0.55*\dlY)$) to node[]{} ($(Leg)+(0.75*\dlX,-0.55*\dlY)$);
    
\end{circuitikz}
}
	\caption
	{%
	    Schematic diagram of the test system, which is based on the \CIGRE low-voltage benchmark microgrid~\cite{Rep:2014:CIGRE} (in black) and extended by unbalanced impedance loads (in grey). For the cable parameters see Table~\ref{tab:grid:parameters}.
	    The resources are composed of constant impedance loads (Z) and constant power loads (P/Q), parameters given in Table~\ref{tab:resources:references}.
	}
	\label{fig:grid:schematic}
\end{figure}
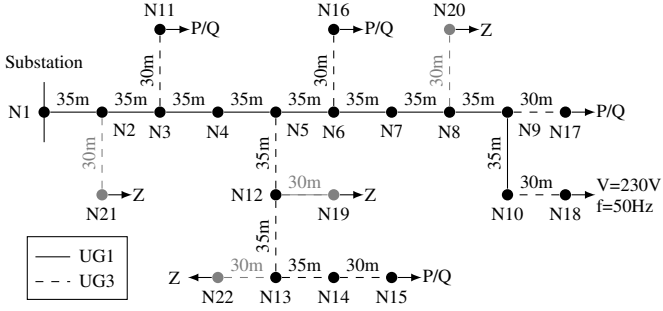

The accuracy of the fixed-point algorithm and the uniqueness of the solution for the \HPF  are verified through numerical analyses on a test system which is adapted from the \CIGRE low-voltage benchmark microgrid~\cite{Rep:2014:CIGRE}. The network's schematic diagram in \cref{fig:grid:schematic} depicts the substation at node N1, five grid-following \CIDER[s] at nodes N11 and N15-N17, one grid-forming \CIDER at node N18 and four unbalanced loads at nodes N19-N22.

The network lines are constructed from underground cables, whose sequence parameters are provided in \cref{tab:grid:parameters}. 

\begin{table}[t]
    \centering
    \caption{Sequence Parameters of the Lines in the Test System.}
    \label{tab:grid:parameters}
	{
\renewcommand{\arraystretch}{1.1}
\setlength{\tabcolsep}{0.15cm}

\begin{tabular}{ccccccc}
    \hline
        ID
    &   $R_{+}/R_{-}$ 
    &   $R_{0}$ 
    &   $L_{+}/L_{-}$ 
    &   $L_{0}$ 
    &   $C_{+}/C_{-}$ 
    &   $C_{0}$ 
    \\
    \hline
        UG1
    &   0.162~$\Omega$
    &   0.529~$\Omega$
    &   0.262~mH
    &   1.185~mH
    &   637~nF
    &   388~nF
    \\
        UG3
    &   0.822~$\Omega$
    &   1.794~$\Omega$
    &   0.270~mH
    &   3.895~mH
    &   637~nF
    &   388~nF
    \\
    \hline
\end{tabular}
}
\end{table}

\begin{table}[t]
    \centering
    \caption{Short-Circuit Parameters of the Th{\'e}venin Equivalent.}
    \label{tab:TE:parameters}
    {

\renewcommand{\arraystretch}{1.1}

\begin{tabular}{cccl}
    \hline
        Parameter
    &   Resource
    &   System
    &   Description
    \\
    &   Validation
    &   Validation
    \\
    \hline
        $V_{n}$
    &   230\,V-\RMS
    &   230\,V-\RMS
    &   Nominal voltage 
    \\
        $S_{\mathit{sc}}$
    &   267\,kW
    &   3.85\,MW
    &   Short-circuit power
    \\
        $\Abs{Z_{\mathit{sc}}}$
    &   195\,m$\Omega$
    &   13.7\,m$\Omega$
    &   Short-circuit impedance
    \\
        $R_{\mathit{sc}}/X_{\mathit{sc}}$
    &   6.207
    &   0.271
    &   Resistance-to-reactance ratio
    \\
    \hline
\end{tabular}
}
\end{table}

\begin{table}[]
	\centering
	\caption{Harmonic Voltages of the Th{\'e}venin Equivalent (see \cite{Std:BSI-EN-50160:2000}).}
	\label{tab:TE:harmonics}
	{
\renewcommand{\arraystretch}{1.1}
\begin{tabular}{ccr}
    \hline
        $h$
    &   $|V_{\TE,h}|$
    &   \multicolumn{1}{c}{$\angle V_{\TE,h}$}
    \\
    \hline
        1
    &   1.0\,p.u.
    &   0\,rad
    \\
        5
    &   6.0\,\%
    &   $\pi$/8\,rad
    \\
        7
    &   5.0\,\%
    &   $\pi$/12\,rad
    \\ 
        11
    &   3.5\,\%
    &   $\pi$/16\,rad
    \\
        13
    &   3.0\,\%
    &   $\pi$/8\,rad
    \\
        17
    &   2.0\%
    &   $\pi$/12\,rad
    \\
        19
    &   1.5\,\%
    &   $\pi$/16\,rad
    \\
        23
    &   1.5\,\%
    &   $\pi$/16\,rad
    \\
    \hline
\end{tabular}
}

\end{table}

\begin{table}[]
    \centering
    \caption{Parameters of the Grid-Following Resources and Loads\newline in the Test System.}
    \label{tab:resources:references}
	{

\renewcommand{\arraystretch}{1.2}
\begin{tabular}{ccccc}
	\hline
		Node
	&	S
	&	pf
	&	Type
	&   Phase weights
	\\
	\hline
		N11
	&	\phantom{-}15.0~kW
	&	0.95
	&   P/Q
	&   [0.33 0.33 0.33]
	\\
		N15
	&	\phantom{-}52.0~kW
	&	0.95
	&   P/Q
	&   [0.33 0.33 0.33]
	\\
		N16
	&	\phantom{-}55.0~kW
	&	0.95
	&   P/Q
	&   [0.33 0.33 0.33]
	\\
		N17
	&	\phantom{-}35.0~kW   
	&	0.95
	&   P/Q
	&   [0.33 0.33 0.33]
	\\
		N19
	&	-51.2~kW   
	&	0.95
	&   Z
	&   [0.31 0.50 0.19]
	\\
		N20
	&	-51.7~kW   
	&	0.95
	&   Z
	&   [0.45 0.23 0.32]
	\\
		N21
	&	-61.5~kW   
	&	0.95
	&   Z
	&   [0.24 0.39 0.37]
	\\
		N22
	&	-61.9~kW   
	&	0.95
	&   Z
	&   [0.31 0.56 0.13]
	\\
	\hline
\end{tabular}
}
\end{table}


The substation is modelled as a \textit{Thévenin Equivalent} (\TE) whose equivalent impedance, encompassing the upstream grid and the substation transformer, is given by the short-circuit parameters in \cref{tab:TE:parameters}. 
Additionally, the \TE equivalent voltage source includes harmonics, whose voltage levels are detailed in  \cref{tab:TE:harmonics}. 
Moreover, the grid-forming \CIDER is assumed to regulate the PCC voltage and frequency at their nominal value (i.e., V=230~V and f = 50~Hz, respectively). 
On the other hand, parameters for the grid-following units and the unbalanced passive loads are detailed in \cref{tab:resources:references}, where the phase weights denote the distribution of the load among the three phases.

\subsection{Validation of the Fixed-Point Algorithm}
\begin{figure}[b]
\psfrag{Iteration}{Iterations}
    \centering
    \includegraphics[width = 0.48\textwidth]{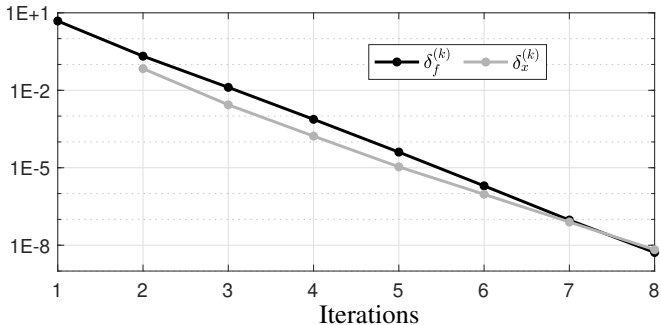}
    \caption{Plot of indicators \( \delta_f^{(k)}\) (in black) and \( \delta_x^{(k)}\) (in grey) across iterations.}
    \label{fig:Residuals}
\end{figure}
\begin{figure}[]
	\centering
    \includegraphics[width = 0.5\textwidth]{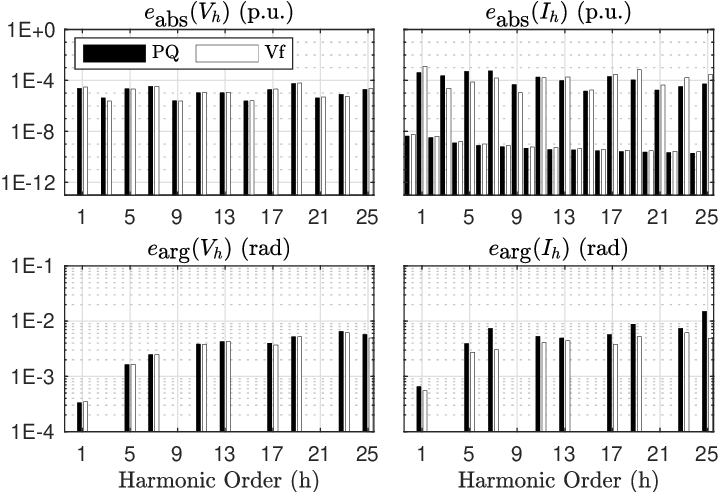}
	\caption
	{%
	    Results of fixed-point algorithm validation on the benchmark system. The plots show maximum absolute errors over all nodes and phases between the time domain simulation and the fixed point algorithm, for voltages (left column) and currents (right column) in magnitude (top row) and phase (bottom row).
	}
	\label{fig:results:errors}
\end{figure}
This section is dedicated to validating the suitability of the proposed fixed-point algorithm for solving the \HPF problem. Particularly, in accordance with~\eqref{eq:convergence}, the convergence of the algorithm relies on the values of two indicators, \(\delta_{x}^{k}\) and  \(\delta_{f}^{k}\), defined as follows:
\begin{alignat}{3}
\delta_{f}^{(k)} =~& \| \boldsymbol{\Phi}(\Tilde{\W}_{\rho}^{(k)})\|_\infty \quad &k\geq 1\\
    \delta_x^{(k)} =~& \| \Tilde{\W}_{\rho}^{(k)}- \Tilde{\W}_{\rho}^{(k-1)}\|_\infty \quad &k\geq 2
\end{alignat}
The former measures the distance between consecutive iterations (i.e., \(k\) and \(k+1\)), whereas the latter quantifies the residual of the function. These indicators serve as stopping criteria. Specifically, the fixed-point iterations stop and converge to the solution when both arbitrarily \(\delta_x^{(k)}\) and \(\delta_f^{(k)}\)  fall below their respective tolerances \(\delta_x^{*}\) and  \(\delta_f^{*}\),  both set at 1E-8. In this regard, Fig.~\ref{fig:Residuals} illustrates the behaviour of these indicators across iterations for the case under study. In particular, the algorithm exhibits linear convergence. More precisely, according to~\cite{quarteroni2010matematica}, it is possible to define the asymptotic convergence factor \(\rho\)  as 
\begin{equation}
    \rho = \log{\left(\|\nabla\boldsymbol{\Phi}(\Tilde{\W}^*_{\rho})\|_{\infty}\right)}
\end{equation}
that in the current case study is equal to -1.675.

Similar to the validation of the \NR method in~\cite{becker2021harmonic_part_II}, the accuracy and performance of the proposed algorithm are assessed through a comparison of the harmonic phasors obtained using the fixed-point algorithm and the \DFT spectra obtained through \textit{Time Domain Simulations} (\TDS) with Simulink. 
In this regard, the same \textit{Key Performance Indicators} (KPIs) are adopted, namely:
\begin{alignat}{2}
    e_{\text{abs}}(\mathbf{X}_h) \coloneqq& \max_{p} \left|\left|X_{h,p,\text{\HPF}}|-|X_{h,p,\text{\TDS}}\right|\right| \label{eq:error:abs}\\
    e_{\text{arg}}(\mathbf{X}_h) \coloneqq& \max_{p} \left|\angle{X_{h,p,\text{\HPF}}}-\angle{X_{h,p,\text{\TDS}}}\right| \label{eq:error:arg}
\end{alignat}
where \(e_{\text{abs}}(\mathbf{X}_h)\) and \(e_{\text{arg}}(\mathbf{X}_h)\) are the maximum absolute errors in magnitude and phase, respectively, over all phases \(p\in \mathcal{P}\) for each Fourier coefficient \(\mathbf{X}_h\) of both voltage and current.

The \TDS are performed in Simulink and the simulations are stopped once the system has reached steady-state conditions. Subsequently, current and voltage spectra are calculated. All analyses in this context are conducted using normalised units w.r.t. the base power \(P_b\)=10~kW and the base voltage  \(V_b\)=230~V-\RMS.

\cref{fig:results:errors} illustrates the maximum absolute errors, as defined in \eqref{eq:error:abs} and \eqref{eq:error:arg} over all nodes and phases, separately for grid-forming and grid-following \CIDER[s]. The highest error w.r.t. voltage magnitude and phase are \(e_{\text{abs}}(\V_{19})=6.33\)E-5~p.u. and \(e_{\text{arg}}(\V_{23})=6.51\)E-3~rad, respectively. Conversely, the highest error w.r.t current magnitude and phase are \(e_{\text{abs}}(\I_{1})=1.33\)E-3~p.u. and \(e_{\text{arg}}(\I_{25})=1.51\)E-2~rad, respectively. Notably, these error levels are generally very low, comparable to the maximum measurement errors of typical instrumentation. Additionally, these findings align with the results obtained through the \NR method reported in~\cite{becker2021harmonic_part_II}.

\subsection{Assessment of the Uniqueness of Solution}
\begin{figure}[]
    \psfrag{title}{\raisebox{0.4em}{$\|\nabla\boldsymbol{\Phi}\|_\infty$}}
    \psfrag{Iterations}{Iterations}
    \centering
    \includegraphics[width = 0.48\textwidth]{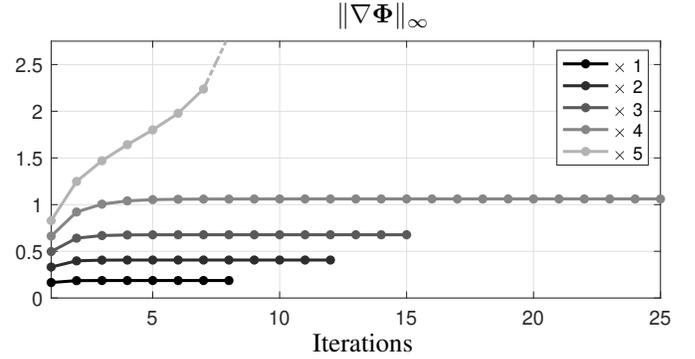} 
    \caption{Norm of the Jacobian of the \(\boldsymbol{\Phi}\) over fixed-point iterations, illustrating the impact of increasing power in P/Q nodes.}
    \label{fig:FixedPoint:Jacobian}
\end{figure}
This section aims at evaluating the uniqueness of the solution, as granted by \cref{Th:1}. According to the theorem, when the fixed-point algorithm converges to a solution, it is considered unique in its neighbourhood if the norm of the Jacobian calculated at that point is smaller than one. 
Specifically, the following explores different scenarios in which the complex power of each grid-following \CIDER has been progressively increased until exceeding the grid loadability. Specifically, the reference power of each grid-following \CIDER has been increased. This affects the entire spectrum of harmonic currents and voltages, as their iso-frequency components contribute to establishing the reference value of power absorbed/injected.

The analysis is conducted on the system described in the preceding section, where the complex power of the \CIDER[s] at nodes N11, N15-N17 has been scaled by a factor ranging from 1 to 5. 
Particularly, \cref{fig:FixedPoint:Jacobian} shows the behaviour of \(\|\nabla\boldsymbol{\Phi}\|_\infty\) over the iterations while varying the scaling factor. 
As the injected power of the \CIDER[s] increases, the asymptotic convergence factor decreases, thereby reducing the algorithm's convergence rate. 
From a physical perspective, this leads to a diminished margin in the harmonic stability of the system. 
Notably, for a scaling factor between 1 and 3, the algorithm successfully identifies unique solutions in their neighborhood, as the Jacobian exhibits a norm smaller than one. 
These solutions can be referred to as stable. 
By contrast, for the scaling factor equal to 4, the algorithm converges, but no definitive conclusion can be drawn about the uniqueness of the solution since the norm of the Jacobian is slightly higher than one. 
Finally, for a scaling factor greater than 5, the system exceeds its loadability, and no solution is found.

\section{Conclusion}
\label{sec:Conclusion}

This paper proposed a new formulation of the \HPF study based on the fixed-point algorithm. 
The algorithm was implemented in MATLAB and validated against \TDS in Simulink. 
The largest observed errors in current and voltage magnitudes were 1.33E-3~p.u. and 6.33E-5~p.u., respectively. 
Similarly, the largest errors in current and voltage phases were 6.51E-3~rad and 1.51E-2~rad, respectively. 
Furthermore, the paper introduced a suitable sufficient condition, based on the Banach-Cacciopoli theorem, for the ex-post certification of the uniqueness of the solution. 
More precisely, once the fixed-point algorithm converges to a solution, assessing the algorithm's contraction property enables the definition of the solution's nature.  In this regard, the approach is validated by verifying the \HPF contraction property through a progressive increase in the reference power absorbed by the grid-following \CIDER[s], eventually exceeding the loadability limits of the network. In such a case the numerical results show that the \HPF problem loses its contractive property, making the problem unsolvable.
\section*{Acknowledgement}
The authors would like to acknowledge Dr. Yihui Zuo, from \textit{LEM International SA}, for her insights and constructive feedback, which contributed to addressing the analytical challenges in this work.
\appendices
\section{}\label{sec:Appendix}
It has to be remarked that the \HPF model presented in~\cite{kettner2021harmonic_part_I}, along with considering \CIDER[s], may consider the potential generation of harmonics caused by various sources, including conventional resources or interactions with upstream and downstream power grids. Additionally, it accounts for the presence of unbalanced passive loads, which are typically connected to distribution networks. In this regard, the fundamentals for the modelling of sources of harmonics other than \CIDER[s] and passive loads are reported below.

\subsection*{Modelling of Sources of Harmonics Other Than CIDERs}
The source of harmonics which are not converter-interfaced can be modelled by transfer functions in the frequency domain. A suitable approach is represented by the application of the \TE or the \textit{Norton Equivalent} (\NE) model. Let $m \in \nodes$ be the node at which the non-\CIDER resource is located, based on the \TE, the injected current $\Hat{\I}_{m}$ is given by
\begin{equation}\label{eq:TheveninEquivalent}
    \Hat{\I}_{m} = \Hat{\Z}_{\TE,m}^{-1}(\Hat{\V}_{m}-\hat{\V}_{\TE,m})
\end{equation}
where $\Hat{\V}_{m}$, $\Hat{\V}_{\TE}$ and $\Hat{\Z}_{\TE}$ are the harmonic voltage of node $m$, the harmonic voltage source and the harmonic impedance, respectively, of the \TE.
If a \NE is used instead, the injected current is given by
\begin{equation}\label{eq:NortonEquivalent}
    \Hat{\I}_{m} = \Hat{\I}_{\NE,m} - \Hat{\Y}_{\NE,m}\Hat{\V}_{m}
\end{equation}
where $\Hat{\I}_{\NE}$ and $\Hat{\Y}_{\NE}$ are the harmonic current source and harmonic admittance, respectively, of the \NE.

To be included in the \HPF framework, these resources can be modelled as either grid-forming or grid-following nodes based on whether  $\Hat{\I}_{m}$ or $\Hat{\V}_{m}$ is chosen as the grid disturbance (see eq.~\eqref{eq:grid:disturbance:time} for clarifications). For instance, according to~\eqref{eq:CIDER:transfer:outer} the grid response of a not-converter-interfaced resource modelled as a grid following \TE  is given by:
\begin{equation}
    \Hat{\I}_{m} = \underbrace{\left[ \begin{array}{cc}
         -\Hat{\Z}_{\TE,m}^{-1} & \Hat{\Z}_{\TE,m}^{-1}
    \end{array}\right]}_{\hat{\G}_m} \left[ \begin{array}{c}
         \Hat{\V}_{m}  \\
         \Hat{\V}_{\TE,m} 
    \end{array}\right]
\end{equation}
where $\hat{\G}_m$ is the equivalent closed-loop gain matrix associated with the \TE resource, $\Hat{\I}_{m}$ is the output vector $\hat{\mathbf{Y}}_{\gamma,m}$, whereas  $\Hat{\V}_{m}$ 
         and $\Hat{\V}_{\TE,m}$ are the grid disturbance $\Hat{\mathbf{W}}_{\gamma,m}$ and the equivalent control-software disturbance vector $\Hat{\mathbf{W}}_{\kappa,m}$, respectively. 
         Other models, such as \TE grid-forming, \NE grid-following or grid following, can be obtained similarly.
         
\subsection{Modelling of Constant Impedance Loads}
Distribution lines usually experience the widespread presence of passive balanced/unbalanced loads, which can be connected among the grid nodes without the neutral conductor or between the nodes of the grid and the neutral conductor. Regardless of their configuration, they are represented by a constant impedance matrix\footnote{It is important to note that both constant current and constant power loads models can be reduced to that of the grid-following \CIDER.}. Passive loads connected among the grid nodes without the neutral conductor can be included in the grid branches primitive compound admittance matrix, as illustrated in~\eqref{eq:branches:primitive}. Passive loads connected between the grid nodes and the neutral can be incorporated into the grid model as shunt elements through the shunts primitive compound admittance matrix. More precisely, let $\passiveloads \subset \mathcal{T}$ denote the subset of nodes where the shunt passive loads are connected, and $\mathbf{Z}_p$ represent their constant three-phase impedance matrix. The associated admittance matrix $\Y_{\passiveloads}(f)$ can defined as follows:
\begin{equation}
    \Y_{\passiveloads}(f)\coloneqq \diag_{p\in \shunts}\left(\Y_{p}(f)\right) \in \mathbb{C}^{|\nodes|\times |\nodes|}
\end{equation}
where $\Y_{p}$ is given by:
\begin{equation}
    \Y_{p}(f) \coloneqq \left\{ \begin{array}{ll}
         \mathbf{Z}_p^{-1}&  \text{if}\  p \in \passiveloads \\
        \mathbb{O} & \text{if}\  p \in \shunts \backslash \passiveloads
    \end{array}\right. \in \mathbb{C}^{3 \times 3}
\end{equation}
Hence, \eqref{eq:nodes:admittance} can be modified to account for the presence of passive loads as follows:
\begin{equation}
    \Y(f)=\mathbf{A}^{T}_{\branchgraph}\Y_{\branches}(f)\mathbf{A}_{\branchgraph} + \Y_{\shunts}(f)+\Y_{\passiveloads}(f)
        \label{eq:nodes:admittanceLoads}
\end{equation}
\bibliographystyle{IEEEtran}
\bibliography{mybib}
\end{document}